\documentclass{elsart}
\usepackage{graphicx}
\usepackage{amssymb}
\usepackage{rotate}
\usepackage{rotating}

\newcommand{\imag}{\mbox{i}}
\newcommand{\iimag}{\mbox{\scriptsize i}}
\newcommand{\expo}{\mbox{e}}
\newcommand{\g}[1]{\mbox{#1}}

\begin{document}

\begin{frontmatter}
\title{Controlling spatiotemporal chaos\\
in oscillatory reaction-diffusion systems\\
by time-delay autosynchronization}

\author{C. Beta},
\author{A. S. Mikhailov\corauthref{cor1}},
\corauth[cor1]{Corresponding author.
Tel.: +49-30-8413-5122;
fax: +49-30-8413-5106.
\ead{mikhailov@fhi-berlin.mpg.de}}

\address{Abteilung Physikalische Chemie,\\
Fritz-Haber-Institut der Max-Plack-Gesellschaft,\\
Faradayweg 4--6, 14195 Berlin, Germany}

\begin{abstract}
Diffusion-induced turbulence in spatially extended oscillatory
media near a supercritical Hopf bifurcation can be
controlled by applying global time-delay autosynchronization.
We consider the complex Ginzburg-Landau equation in the Benjamin-Feir
unstable regime and
analytically investigate the stability of uniform oscillations
depending on the feedback parameters.
We show that a noninvasive stabilization of uniform oscillations
is not possible in this type of systems.
The synchronization diagram in the plane spanned by the
feedback parameters is derived.
Numerical simulations confirm the analytical results and
give additional information on the spatiotemporal dynamics of the
system close to complete synchronization.
\end{abstract}

\begin{keyword}
Reaction-diffusion systems \sep Turbulence \sep Feedback

\PACS 05.45.+b \sep 82.40.Bj \sep 82.20.Wt
\end{keyword}
\end{frontmatter}
%
\section{Introduction}
\label{sc:introduction}
Pattern formation in spatially extended systems far from thermal
equilibrium has attracted much attention in recent years and became
a field of active research
\cite{cross-hohenberg-1993,mikhailov-book1}.
A large class of these systems is described by equations of the
reaction-diffusion type.
In such systems, a broad variety of complex
spatiotemporal patterns has been observed, ranging from
travelling pulses and target patterns to rotating spiral waves and
spatiotemporal chaos.
Important examples are chemical reaction-diffusion
systems such as the Belousov-Zhabotinsky reaction
or catalytic surface chemical reactions
\cite{zaikin-zhabotinsky-1970,winfree-1972,imbihl-ertl-1995},
as well as semiconductor and biological systems
\cite{schoell-book-2001,benjacob-cohen-levine-2000}.

Many systems show oscillatory dynamics of their individual elements.
Close to the onset of oscillations, the equations describing
any oscillatory system can be reduced
to a simple universal equation for the complex oscillation amplitude, the
{\itshape complex Ginzburg-Landau equation} (CGLE)
\cite{kuramoto-book,aranson-kramer-2002}.
For a distributed system with diffusive coupling, the CGLE corresponds to
the normal form of a supercritical Hopf bifurcation.
Although the CGLE is strictly valid only
in the immediate vicinity of the oscillation onset, it has been found in
many cases that its qualitative predictions also hold within a wider range
near the bifurcation point.
The CGLE thus provides a general framework for studying common features
of the dynamical behaviour in oscillatory reaction-diffusion systems.
If oscillations are synchronized, uniform oscillations can be
established in the medium or regular spatiotemporal patterns
can emerge.

If the Benjamin-Feir criterion is met, synchronous oscillations become
unstable and a turbulent state develops
where two types of chaotic behaviour can be distinguished.
In the state of phase turbulence,
the local oscillation phase exhibits weak irregular fluctuations while
the real amplitude remains saturated showing only small variations
\cite{kuramoto-book}.
Amplitude or defect turbulence
is characterized by strong fluctuations of both phase and amplitude
which are due to the presence of defects, the locations with a vanishing
amplitude
\cite{coullet-gil-lega-1989}.

Over the past decade,
problems of chaos control became to play a central role in the
studies of chaotic dynamics
\cite{schuster-book-1999}.
Inspired by the pioneering work of Ott, Grebogi, and Yorke (OGY)
\cite{ott-grebogi-yorke-1990},
control of chaos has been studied in the context of many different
dynamical systems
\cite{ditto-rauseo-spano-1990,carroll-et-al-1992,petrov-et-al-1993}
However, the OGY method is designed to control chaos in
low-dimensional dynamical systems.
It requires continuous extensive analysis of the system dynamics
which is virtually impossible to carry out in the case of fully developed
high-dimensional turbulence in spatially extended systems.
A different empirical control method based on implementing a delayed
feedback loop was proposed by Pyragas
and is known as {\itshape time-delay autosynchronization}
(TDAS)~\cite{pyragas-1992,socolar-sukow-gauthier-1994}.
In this method, a feedback signal $F$ is applied to the system that is
proportional to the
difference between the delayed value of a given system variable $u$ and
its instantaneous value,
$F \sim u(t-\tau) - u(t)$.
This approach is easily extended to spatially distributed experimental
systems that often do not allow individual addressing of their local
elements.
Then, the global feedback signal is generated from the integral
value of $u$ over all system elements.
In recent studies, the TDAS method has been used in a number of
different dynamical systems
\cite{pierre-bonhomme-atipo-1996,franceschini-bose-schoell-1999,parmananda-et-al-1999}
and was modified and improved considerably
\cite{beck-et-al-2002,baba-et-al-2002,just-et-al-2003}.
For the CGLE, the action of local TDAS has been previously
considered
\cite{harrington-socolar-2001,montgomery-silber-2003pre}.

The present work was motivated by our experimental study of catalytic CO
oxidation under TDAS
\cite{beta-et-al-2003}.
The catalytic CO oxidation on a platinum (110) single crystal surface
constitutes the most thoroughly studied example among a class of simple
heterogeneous catalytic reactions for which reaction-induced
spatiotemporal pattern formation has been reported
\cite{imbihl-ertl-1995}.
It displays a particularly rich variety of spatiotemporal concentration
patterns
\cite{jakubith-et-al-1990}
that can be modeled numerically using a simple model of three coupled
partial differential equations
\cite{krischer-eiswirth-ertl-1992,baer-et-al-1992}.
The effect of global delayed feedback on pattern formation in catalytic
CO oxidation was studied previously using a different time-delay feedback
scheme
\cite{kim-et-al-2001,bertram-et-al-2003,bertram-mikhailov-2003}.
The same feedback scheme was also investigated in detail in the context
of the complex Ginzburg-Landau equation
\cite{battogtokh-mikhailov-1996,battogtokh-preusser-mikhailov-1997}.
Our work on suppression of chemical turbulence in catalytic CO
oxidation using the TDAS scheme
\cite{beta-et-al-2003}
was focused on investigating the question of invasiveness.
We found in the experiment that, although we could reduce the magnitude
of the feedback signal considerably, we were not able to reach the ideal
limit of a vanishing feedback signal in the state of control.
In the present work, we use the complex Ginzburg-Landau equation as a
general model for spatially extended oscillatory systems to study common
aspects of control of diffusion-induced chemical turbulence by TDAS.

The paper is organized as follows. In the next section, the physical
problem and the corresponding equations will be introduced.
In Section \ref{sc:stability}, we analyze the stability of uniform
oscillations in the presence of TDAS and derive a synchronization
diagram.
Numerical simulations of the complex Ginzburg-Landau
equation with TDAS are described in Section
\ref{sc:numerics}.
The paper ends with conclusions and a discussion of the obtained results.
%
\section{Formulation of the problem}
\label{sc:equation}
In absence of feedback, the dynamics of a chemical
reaction-diffusion system can be generally described by a set of
coupled partial differential equations
\begin{equation}
\label{eq:rds}
   \mathbf{\dot{u}}=\mathbf{f}(\mathbf{u}\,,\mathbf{p})+
   \mathbf{D}\nabla^2\,\mathbf{u}\, ,
\end{equation}
where $\mathbf{u}=(u_1,u_2,...,u_N)$ represents concentrations of
reacting species and $\mathbf{D}$ is their diffusion matrix.
The set of nonlinear functions $\mathbf{f}(\mathbf{u},\mathbf{p})$ of
the concentrations $\mathbf{u}$ accounts for the reaction part of the
dynamics and depends on a number of parameters
$\mathbf{p}=(p_1,p_2,...,p_M)$, such as rate constants and external
conditions.

Suppose that the dynamics of system (\ref{eq:rds}) is such that a
supercritical Hopf bifurcation occurs when crossing a certain threshold
$\mathbf{p}=\mathbf{p}_0$ in the parameter space.
Close to the onset of oscillations, system (\ref{eq:rds}) can be
transformed to a simple description of the dynamics in terms of a complex
oscillation amplitude $\eta(x,t)$.
This transformation is based on retaining only the leading critical
modes that govern the dynamics close to the bifurcation point.
It yields as equation of motion for the amplitude $\eta(x,t)$, the complex
Ginzburg-Landau equation
\cite{kuramoto-book},
\begin{equation}
\label{eq:cgle}
   \dot{\eta}=(1-\imag\omega)\eta-(1+\imag\beta)|\eta|^2\eta+
   (1+\imag\varepsilon)\nabla^2\eta \, ,
\end{equation}
where the parameters $\omega$, $\beta$, and $\varepsilon$ denote the
linear oscillation frequency, the nonlinear frequency
shift, and the linear dispersion coefficient, respectively.
When the condition $1+\varepsilon\beta<0$ of the Benjamin-Feir
instability is satisfied, uniform oscillations are unstable and
turbulence spontaneously develops in the system.
To control the turbulence, global feedback can be introduced.
In the presence of global time-delayed feedback, the considered
system is described by the equation
\begin{equation}
\label{eq:cglefeed}
   \dot{\eta}=(1-\imag\omega)\eta-(1+\imag\beta)|\eta|^2\eta+
   (1+\imag\varepsilon)\nabla^2\eta+F(t) \, ,
\end{equation}
where the feedback term is given by
\begin{equation}
\label{eq:fterm}
   F(t)=\mu \, \expo^{\iimag\chi}(\bar{\eta}(t-\tau)-\bar{\eta}(t)) \, ,
\end{equation}
and
\begin{equation}
\label{eq:avampl}
   \bar{\eta}(t)=\frac{1}{L}\int_0^L\eta(x,t)\g{d}x \, .
\end{equation}
Here, $\mu$ is the feedback intensity factor, $\tau$ is the delay time,
and $\chi$ is a phase shift in the application of the control force.

\begin{figure}[htbh]
\begin{center}
\includegraphics[width=10cm]{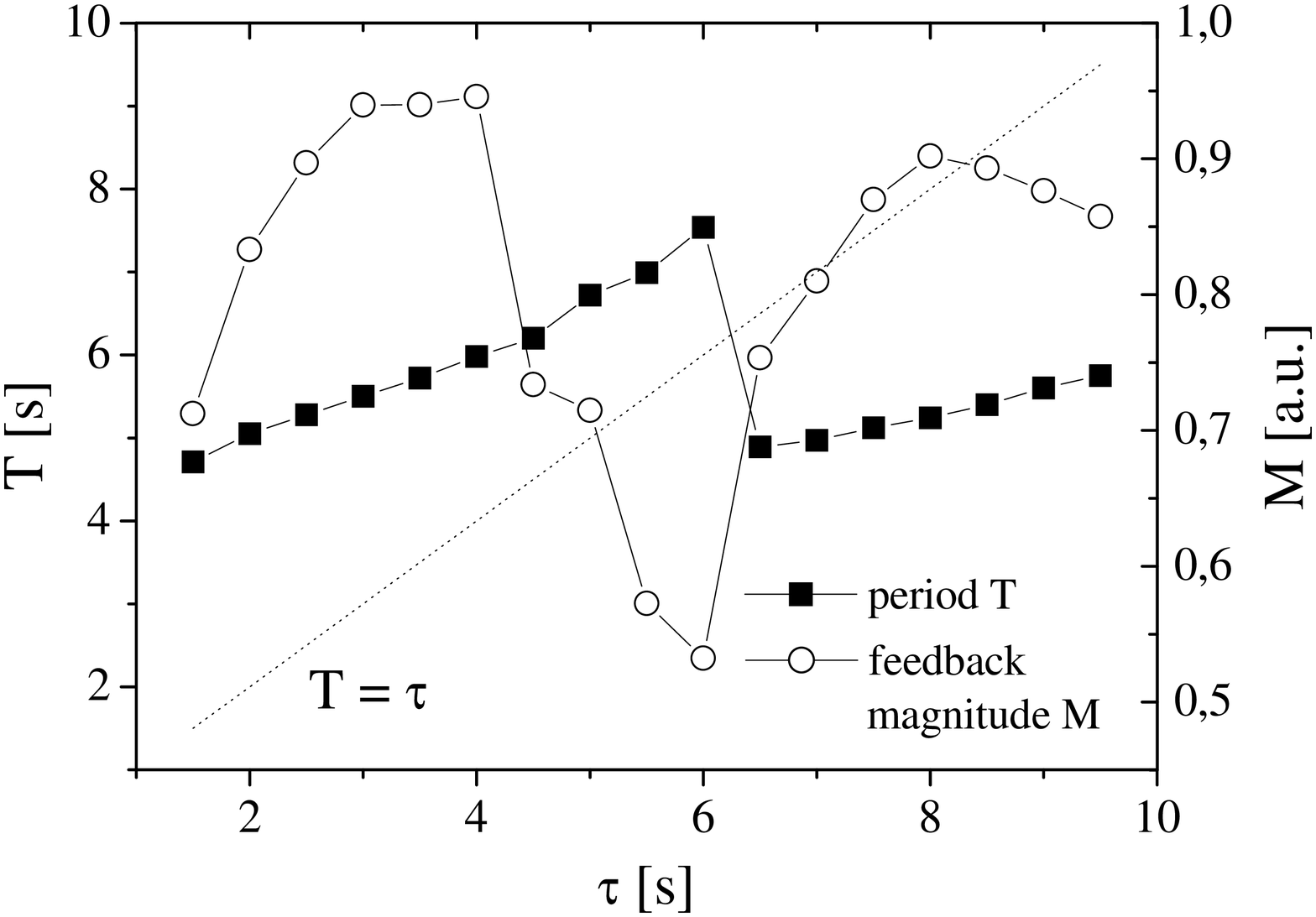}
\end{center}
\caption{Experimental results on the application of TDAS to the
catalytic CO oxidation on Pt(110).
Period $T$ of homogeneous oscillations (black squares) and feedback
magnitude $M$ (open circles) in dependence on the delay time $\tau$.
Reproduced from Ref.~\cite{beta-et-al-2003}.
}
\label{fg:exp}
\end{figure}

Previously,
we showed experimentally~\cite{beta-et-al-2003} that diffusion-induced
turbulence could be
efficiently suppressed in an oscillatory chemical reaction-diffusion
system using TDAS.
It was found that the invasiveness of the control scheme could be significantly
reduced for an optimized choice of feedback parameters.
Figure~\ref{fg:exp} presents the period of uniform oscillations in the state
of control and the feedback magnitude as a function of the delay time
in the experiments with CO oxidation on Pt(110).
Although the feedback magnitude was significantly reduced for a delay
close to $\tau=6$ s, the optimal case of a vanishingly small feedback signal
could not be reached.
Near the point, for which the period $T$ of stabilized uniform
oscillations becomes equal to the delay time $\tau$,
an instability was found.
As seen in Fig.~\ref{fg:exp}, the system avoids a state corresponding
to the intersection point with the line for which $T=\tau$.
The stability of this state, for which the feedback vanishes, has been
earlier investigated~\cite{beta-et-al-2003} for the uniform system
in terms of a phase dynamics equation of a single oscillator.
In this article, we extend the analysis to the theoretical study of
a spatially extended system.

Since our present work was motivated by the experimental results in
\cite{beta-et-al-2003}, we restrict our investigation to the case of a
globally applied control signal generated from the averaged complex
amplitude.
This case is relevant in many experimental situations where a controlled
manipulation of the individual system elements is not possible.
Note, however, that a different behaviour can be expected for a
space dependent application of the control scheme.
%
\section{Linear stability analysis}
\label{sc:stability}
We choose a superposition of a homogeneous mode
$H$ with a small spatially inhomogeneous perturbation
of arbitrary non-zero wave number
$\kappa$ as an ansatz for the complex oscillation amplitude,
\begin{equation}
\label{eq:ansatz_ampl}
   \eta(x,t)=H(t)+A_+(t)e^{\iimag\kappa x}+A_-(t)e^{-\iimag\kappa x} \, .
\end{equation}
Substituting expression (\ref{eq:ansatz_ampl}) into Eq.
(\ref{eq:cglefeed}), we can separate, for small $A_{\pm}$,
homogeneous contributions from the spatially
inhomogeneous terms. 
In this way, we find an equation for the homogeneous mode which is
decoupled from the nonuniform contributions,
\begin{equation}
\label{eq:h}
   \dot{H}=(1-\imag\omega)H-(1+\imag\beta)|H|^2H+
   \mu e^{\iimag\chi}(H(t-\tau)-H(t)) \, ,
\end{equation}
and a pair of coupled equations for the wave amplitudes $A_{\pm}$,
\begin{eqnarray}
   \dot{A}_+&=(1-\imag\omega)A_+-(1+\imag\varepsilon)\kappa^2A_+-
   2(1+&\imag\beta)|H|^2A_+-(1+\imag\beta)H^2A_-^*,\label{eq:a+all} \\
   \dot{A}_-^*&=(1+\imag\omega)A_-^*-(1-\imag\varepsilon)\kappa^2A_-^*-
   2(1-&\imag\beta)|H|^2A_-^*-(1-\imag\beta)H^{*2}A_+.\label{eq:a-all}
\end{eqnarray}
Let us first discuss the dynamics of the uniform mode
described by Eq.~(\ref{eq:h}).
Substituting $H=\rho_0\,e^{-\iimag\Omega t}$ into
Eq.~(\ref{eq:h}) we obtain the amplitude of uniform
oscillations in the presence of TDAS,
\begin{equation}
   \label{eq:rho}
   \rho_0=\sqrt{1+\mu(\cos(\chi+\Omega\tau)-\cos\chi)} \, ,
\end{equation}
and derive the equation
\begin{equation}
   \label{eq:omega}
   \Omega=\omega+\beta+\mu\beta(\cos(\chi+\Omega\tau)-\cos\chi)
   -\mu(\sin(\chi+\Omega\tau)-\sin\chi)
\end{equation}
determining the frequency $\Omega$ of uniform oscillations.
Note that without the feedback ($\mu=0$) the system shows uniform
oscillations with the frequency $\Omega_0=\omega+\beta$.

\begin{figure}[htbh]
\setlength{\unitlength}{1mm}
\begin{center}
\begin{picture}(160,60)
\put(65,45){a}
\put(145,45){b}
\put(80,0){\includegraphics[width=7.5cm]{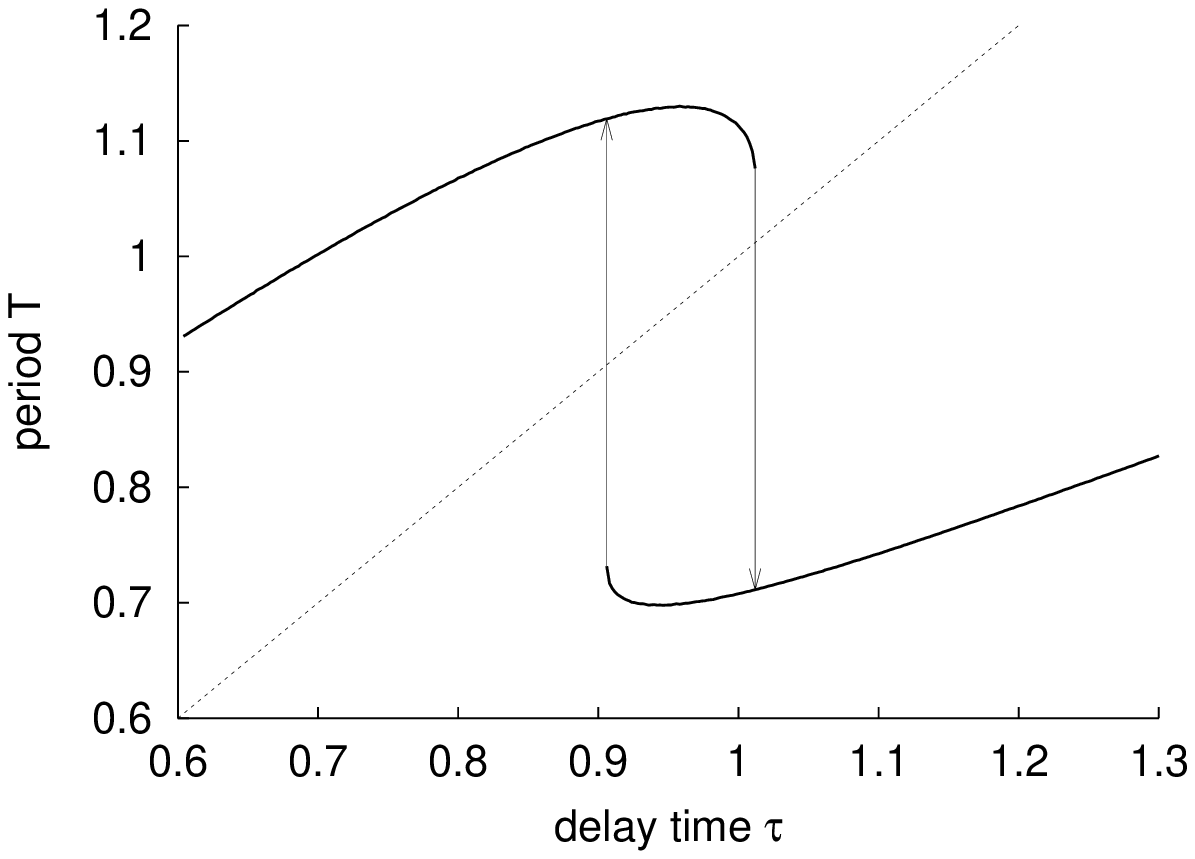}}
\put(0,0){\includegraphics[width=7.5cm]{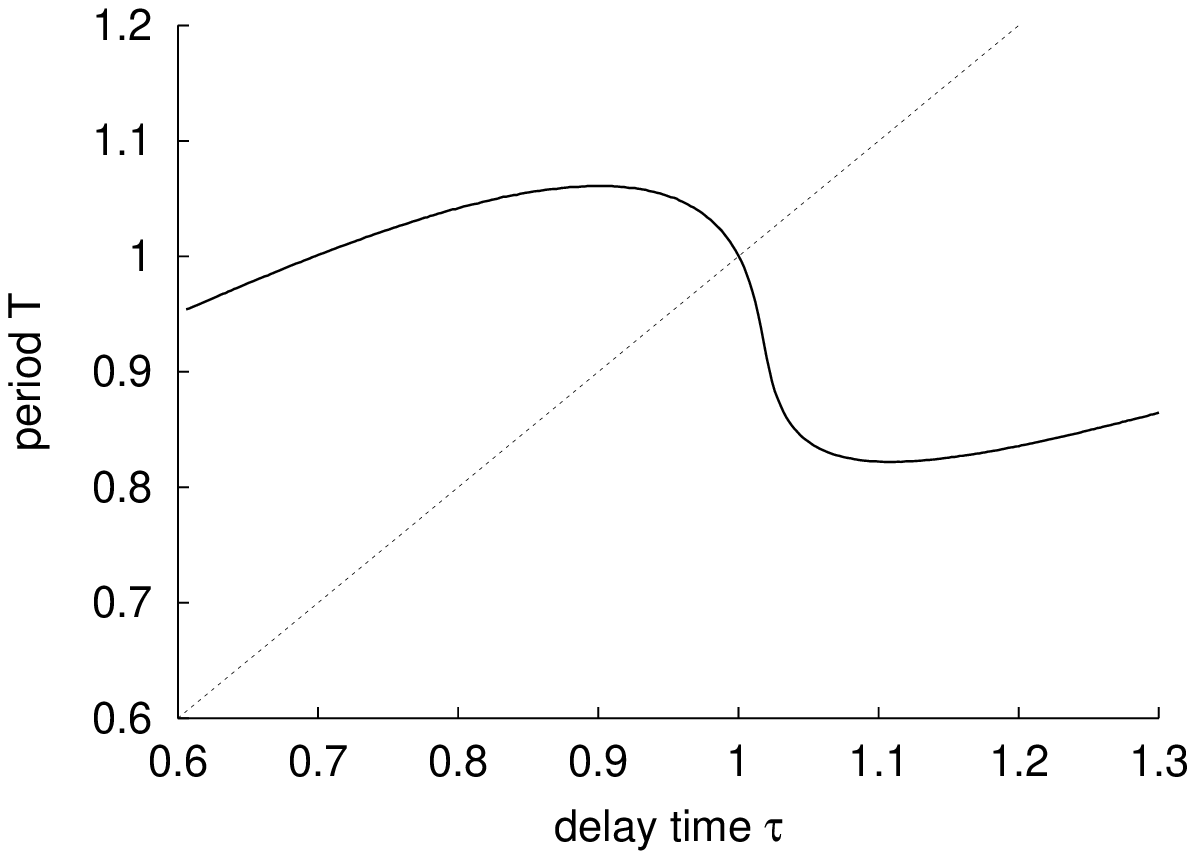}}
\end{picture}
\end{center}
\caption{Results of numerical integration of~Eq.(\ref{eq:h}).
The oscillation period is plotted as a function of the delay time
for (a) low ($\mu=0.5$) and (b) high ($\mu=1.0$) feedback intensities.
The parameters are
$\beta=-1.4$,
$\omega=2\pi-\beta\approx7.68$, and
$\chi=\pi/2$.}
\label{fg:period}
\end{figure}

Figure~\ref{fg:period} shows the results of numerical integration
of Eq.~(\ref{eq:h}).
The oscillation period $T=2\pi/\Omega$ is shown as a function of the
delay time $\tau$ for weak and strong feedbacks.
We see that the state with $T=\tau$ is stable for a weak
feedback and becomes unstable for a high feedback intensity.
The strong feedback case [Fig.~\ref{fg:period}(b)] shows a
good qualitative agreement with the experimental data in
Fig.~\ref{fg:exp}.

\begin{figure}[htbh]
\begin{center}
\includegraphics[width=7cm,angle=-90]{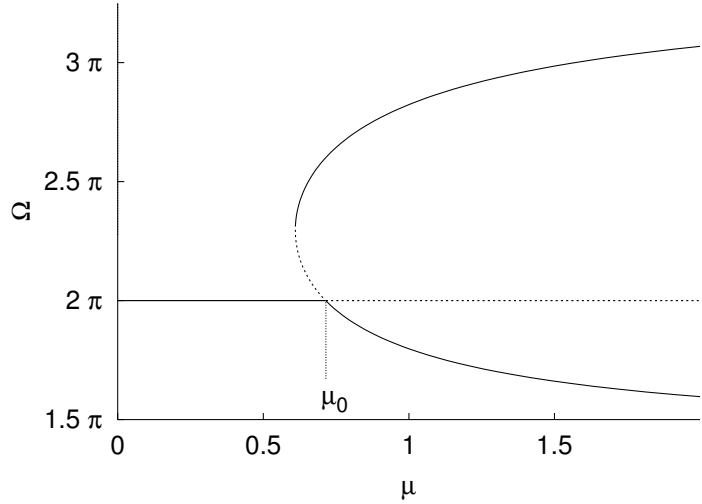}
\end{center}
\caption{Bifurcation diagram for $\tau=T_0=1$.
The parameters are as in Fig. \ref{fg:period}.
Dotted lines denote unstable branches.
The uniform solution with $\Omega=\Omega_0=2\pi$ and
a vanishing feedback signal undergoes a transcritical
bifurcation at $\mu=\mu_0$ and becomes unstable.}
\label{fg:bifur}
\end{figure}

The stability of the state with $T=\tau$ can be understood in
terms of the bifurcation diagram presented in Fig.~\ref{fg:bifur}.
This diagram is constructed by solving Eq.~(\ref{eq:omega})
for a delay time $\tau$ equal to the period $T_0=2\pi/(\omega+\beta)$
of oscillations in the nonperturbed uniform system.
Besides a solution with a vanishing feedback term, for which
$\Omega=\Omega_0$, other solutions with a non-zero feedback and
$\Omega\ne\Omega_0$ are yielded by Eq.~(\ref{eq:omega}).
Linear stability analysis shows that the solution with
$\Omega=\Omega_0$ and a vanishing feedback is stable for small $\mu$ and
becomes unstable beyond some critical feedback intensity $\mu_0$.
These results are similar to those obtained by
using the phase dynamics approximation for a single
oscillator in the presence of TDAS~\cite{beta-et-al-2003}.

We now turn to the stability analysis of uniform oscillations
with respect to spatially inhomogeneous perturbations.
The solution of the linear equations (\ref{eq:a+all}) and (\ref{eq:a-all})
can be sought as
$A_+=A_+^0e^{-i\Omega t}e^{\lambda t}$ and
$A_-^*=A_-^{*0}e^{i\Omega t}e^{\lambda t}$.
Substituting this into (\ref{eq:a+all}) and (\ref{eq:a-all}),
we obtain the following expression
\begin{equation}
\label{eq:lambda}
   \lambda_{1,2}=1-\kappa^2-2\rho_0^2\pm\sqrt{(1+\beta^2)\rho_0^4
   -(\Omega-\omega-\varepsilon\kappa^2-2\beta\rho_0^2)^2} \, ,
\end{equation}
where $\rho_0$ is given by Eq.~(\ref{eq:rho}) and $\Omega$ is
a solution of Eq.~(\ref{eq:omega}).
Uniform oscillations are stable with respect to the growth of
spatially nonuniform modes if Re $\lambda_{1,2}<0$ for all
wavenumbers $\kappa$.
The instability boundary is therefore determined by the
conditions
\begin{eqnarray}
   \mbox{Re}\,\lambda(\mu_c,\kappa_c)&=&0 \, , \label{eq:cond1} \\
   \frac{\partial}{\partial \kappa}\mbox{Re}\,
   \lambda(\mu_c,\kappa_c)&=&0 \, . \label{eq:cond2}
\end{eqnarray}
Here, Eq. (\ref{eq:cond2}) accounts for the fact that the
wave number of the first unstable mode corresponds to a maximum in
$\lambda$ plotted as a function of $\kappa$.
The equations (\ref{eq:cond1}) and (\ref{eq:cond2})
can be solved numerically, taking into account 
Eq. (\ref{eq:omega}) yielding the dependence of the frequency $\Omega$
of the uniform mode on the feedback intensity $\mu$ and the
delay time $\tau$.

We are particularly interested in the stability of uniform
oscillations with $\Omega=\Omega_0$ and $\tau=T_0$.
Inserting this into the
general expression (\ref{eq:lambda}) for $\lambda$, we find
\begin{equation}
\label{eq:lambdaspecial}
   \lambda_{1,2}=-\kappa^2-1\pm
   \sqrt{-\varepsilon^2\kappa^4-2\beta\varepsilon\kappa^2+1} \, ,
\end{equation}
which is independent of $\mu$.
Therefore, this state is always unstable when the Benjamin-Feir
condition $1+\varepsilon\beta<0$ is fulfilled.
All inhomogeneous modes with a wavenumber less than
\begin{equation}
\label{eq:kappaspecial}
   \kappa=\sqrt{-\frac{2(1+\varepsilon\beta)}{1+\varepsilon^2}}
\end{equation}
are growing, no matter how $\mu$ is chosen.
Thus, for $1+\varepsilon\beta<0$ the solution with
$\Omega=\omega+\beta$ will be always unstable so that a noninvasive
stabilization of uniform oscillations with TDAS is not possible in this
type of system.

\begin{figure}
\begin{center}
\includegraphics[width=10cm]{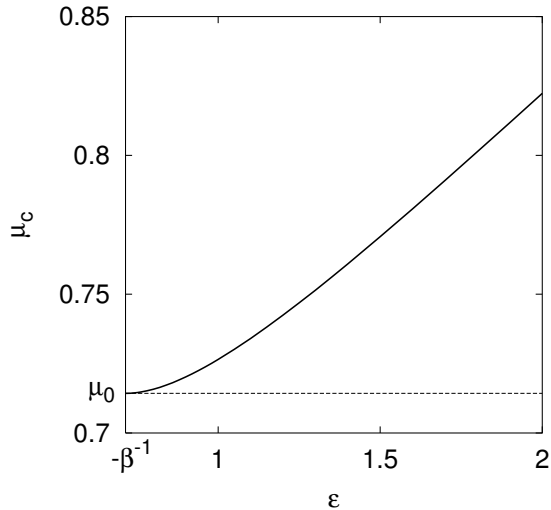}
\end{center}
\caption{Critical feedback intensity $\mu_c$ for $\tau=T_0=1$
as a function of
the dispersion coefficient $\varepsilon$ with $\beta=-1.4$.
The line $\mu=\mu_0$ denotes the feedback intensity for which the
transcritical bifurcation occurs in the uniform system (cf.
Fig.~\ref{fg:bifur}).
The other parameters are as in Fig.~\ref{fg:period}.}
\label{fg:mueps}
\end{figure}

For the other solutions with $\Omega\ne\Omega_0$ at $\tau=T_0$, presented in
Fig.~\ref{fg:bifur}, the feedback signal is nonzero.
In this case, we can determine $\mu_c$ from the general
expression (\ref{eq:lambda}) for $\lambda$ by numerically solving
equations (\ref{eq:cond1}) and (\ref{eq:cond2}).
Figure~\ref{fg:mueps} shows the critical feedback intensity $\mu_c$ as
a function of the dispersion parameter $\varepsilon$ for
$\tau=T_0$ and the other parameters chosen as in Fig.~\ref{fg:bifur}.
In the Benjamin-Feir unstable regime, $\mu_c$ lies above the bifurcation
point $\mu_0$ at which the solution with $\Omega=\Omega_0$
becomes unstable, cf. Fig. \ref{fg:bifur}.
As the system approaches the Benjamin-Feir line with decreasing
$\varepsilon$, the critical feedback intensity necessary to stabilize uniform
oscillations decreases and finally converges towards $\mu_0$.

\begin{figure}[htbh]
\setlength{\unitlength}{1mm}
\begin{center}
\begin{picture}(110,110)
\put(2,57){\includegraphics[width=10cm]{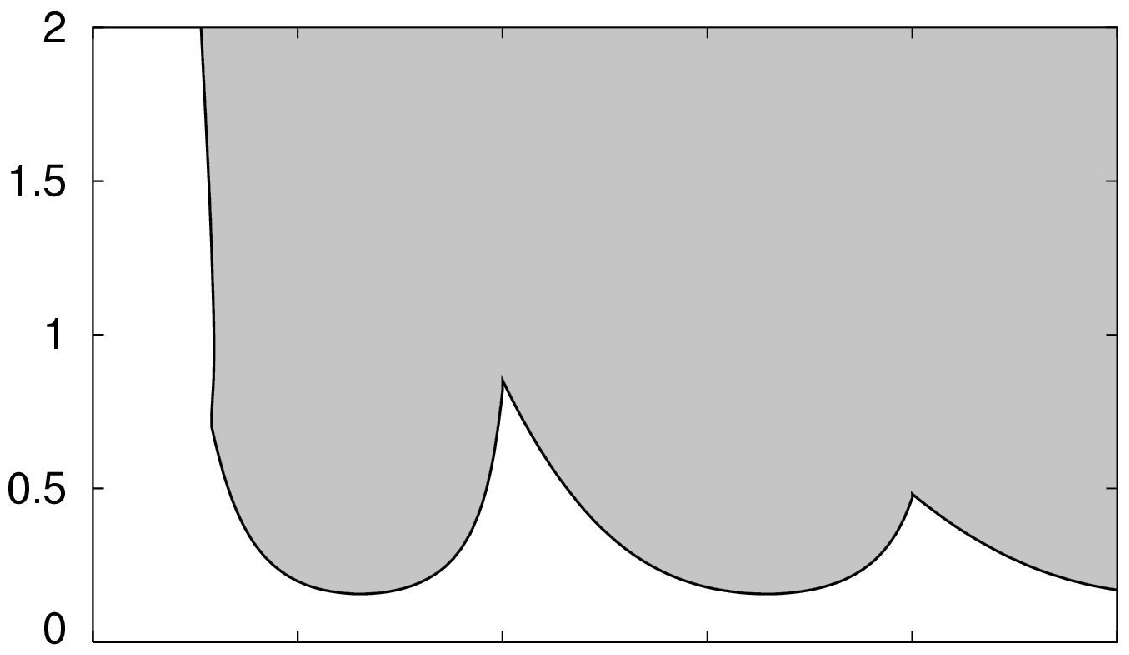}}
\put(91,108){a}
\put(91,56.5){b}
\put(91,29){c}
\put(19,81){A}
\put(45,87){B}
\put(78,78){C}
\put(0,89){$\mu_c$}
\put(0,49){$\kappa_c$}
\put(2,30){\includegraphics[width=10cm]{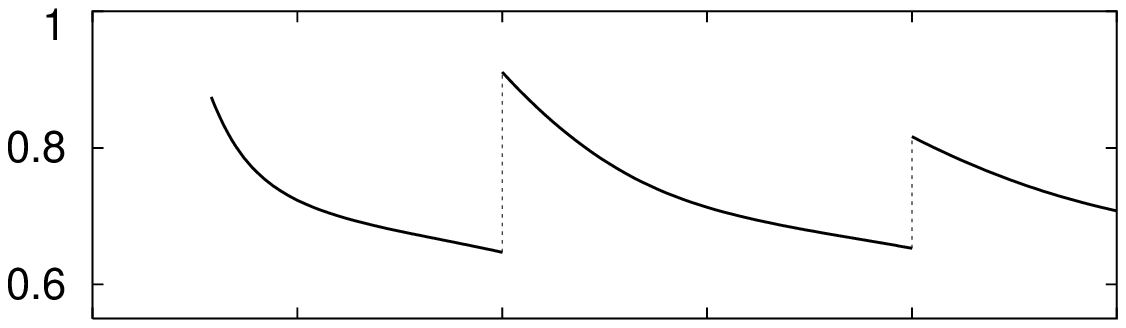}}
\put(0,22){$\Omega_c$}
\put(6.4,3){\includegraphics[width=9.53cm]{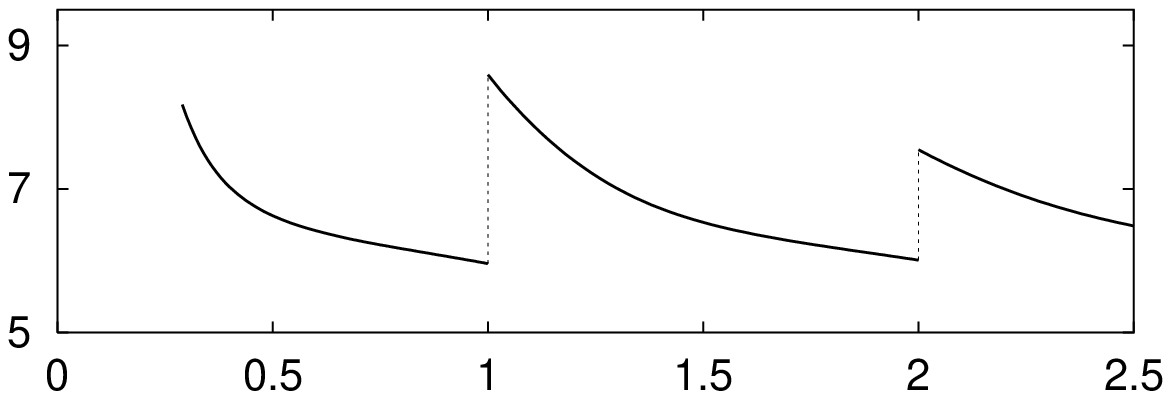}}
\put(54,0){$\tau$}
\end{picture}
\end{center}
\caption{(a) Synchronization diagram.
Uniform oscillations are stable inside the shaded region.
(b,c) The dependences of the critical wavenumber $\kappa_c$ and the
critical frequency $\Omega_c$ on the delay time $\tau$.
The parameters are 
$\varepsilon=2$, 
$\beta=-1.4$,
$\omega=2\pi-\beta\approx7.68$, and
$\chi=\pi/2$.
}
\label{fg:diagram}
\end{figure}

\begin{figure}
\begin{center}
\includegraphics[width=10cm]{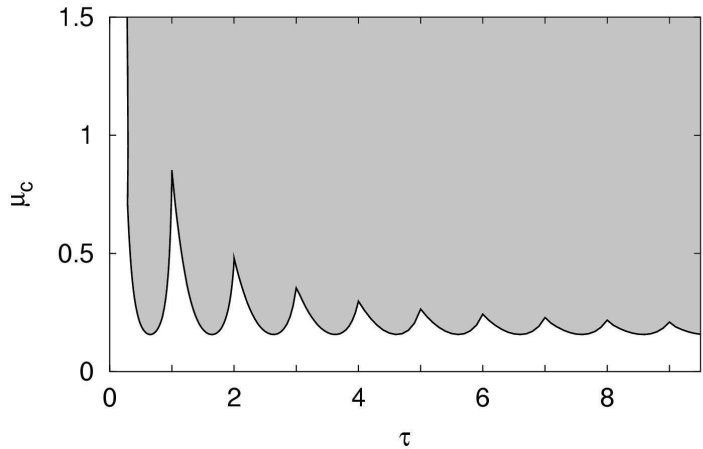}
\end{center}
\caption{Extended synchronization diagram.
The same parameters as in Fig.~\ref{fg:diagram}.}
\label{fg:diagram_large}
\end{figure}

From the conditions (\ref{eq:cond1}) and (\ref{eq:cond2}), we can
also determine the critical feedback intensity $\mu_c$ as a function
of the delay time $\tau$.
The resulting synchronization diagram in the plane spanned by the feedback
parameters $\tau$ and $\mu$ is displayed in Fig.~\ref{fg:diagram}(a).
The curve for the critical feedback intensity $\mu_c$ divides the 
plane into a shaded region, where uniform oscillations are linearly stable
with respect to small perturbations of arbitrary wavenumber, and a region
where uniform osciallations are unstable.
The characteristic feature of this diagram is the repeated
appearance of cusps.
They are observed whenever $\tau$ becomes equal to an integer multiple
of the period of the unperturbed uniform system,
$\tau=k\,2\pi/(\omega+\beta),k=1,2,3,...$ .
Figure~\ref{fg:diagram_large} shows an extension of the top part of
Fig.~\ref{fg:diagram} towards large delay times.
With increasing $\tau$, the cusps get less pronounced and the
boundary converges to a flat line at $\mu\approx0.16$.

According to this analytically derived synchronization diagram, stability
of uniform oscillations can also be maintained by applying global feedbacks
with very large delay times.
Moreover, the critical feedback strength, needed to maintain 
synchronization, does not depend on the delay in the limit
$\tau \rightarrow \infty$.
To understand this result, we note 
that the feedback signal $F(t)$ for $\tau \rightarrow \infty$
is given by
\begin{equation}
\label{eq:taulimit}
   F(t)=\mu e^{i\chi }\left[\rho_{0}e^{-i(\Omega t+\phi_{0})}
        -\bar{\eta}(t)\right]
\end{equation}
where $\phi_{0}$ is a constant phase shift.
The first term here corresponds to the state 
$\bar{\eta}(t-\tau)$ at $\tau \rightarrow \infty$ 
which is essentially the initial state of the system.
When destabilization of initally uniform oscillations is considered to 
determine the stability boundary, this initial state represents uniform 
oscillations with frequency $\Omega$ and amplitude $\rho_{0}$.
Substituting this expression for $F(t)$ into Eq.~(\ref{eq:cglefeed}),
we see that then a situation with external periodic forcing is effectively 
realized.
The critical value $\mu_{c}$ corresponds in this case to the 
minimum forcing intensity needed to maintain uniform oscillations in 
the system.

In Fig.~\ref{fg:diagram}(b,c), the critical
wavenumber and the frequency are shown, respectively, as functions of
the delay time along the lower part ABC of the stability boundary.
The two curves are of similar shape:
they display a decrease for increasing delay time interrupted
by discontinuous jumps.
These discontinuities occur at the locations where the cusps are
found in the stability boundary in Fig.~\ref{fg:diagram}(a).

\begin{figure}[htbh]
\setlength{\unitlength}{1mm}
\hspace{-1.9cm}
\begin{picture}(120,60)
\put(50,43){a}
\put(110,43){b}
\put(170,43){c}
\put(0,0){\includegraphics[width=8cm]{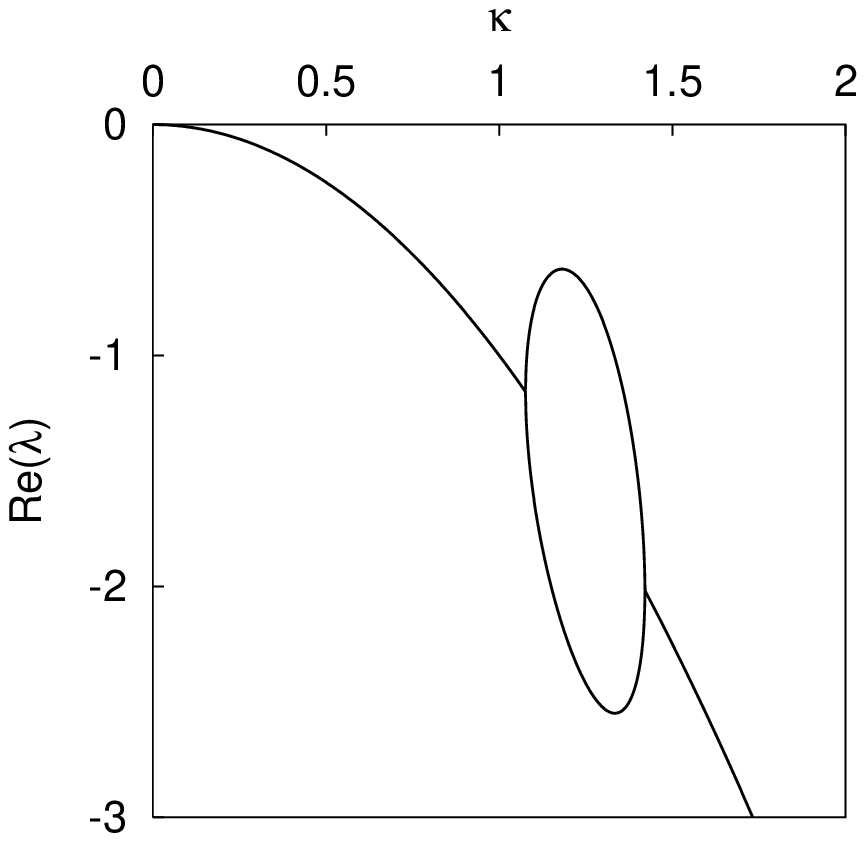}}
\put(60,0){\includegraphics[width=8cm]{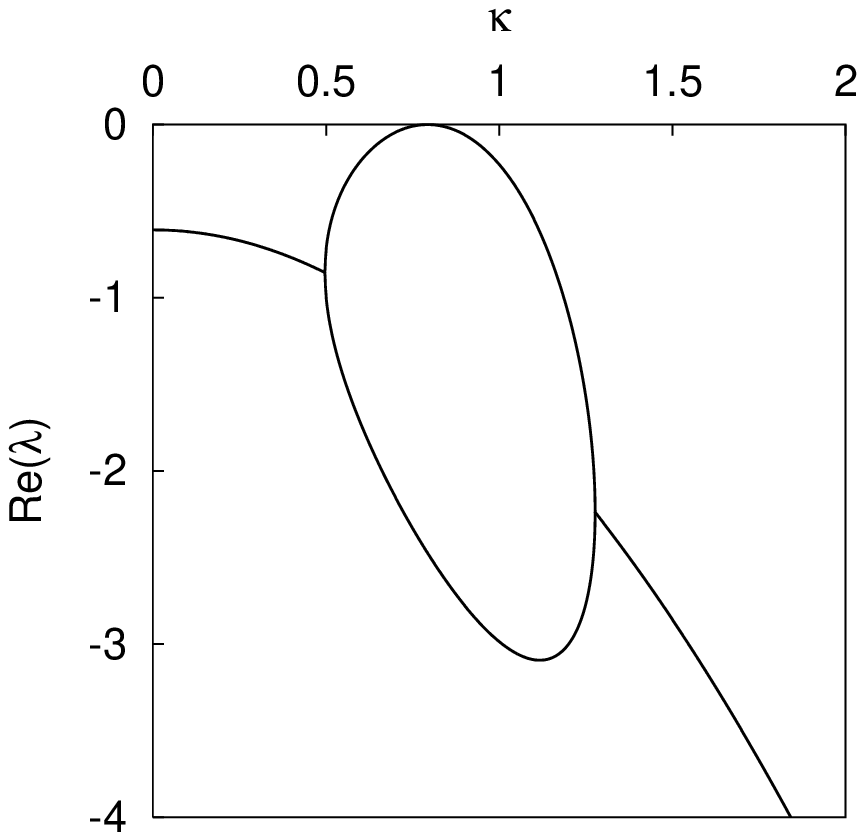}}
\put(120,0){\includegraphics[width=8cm]{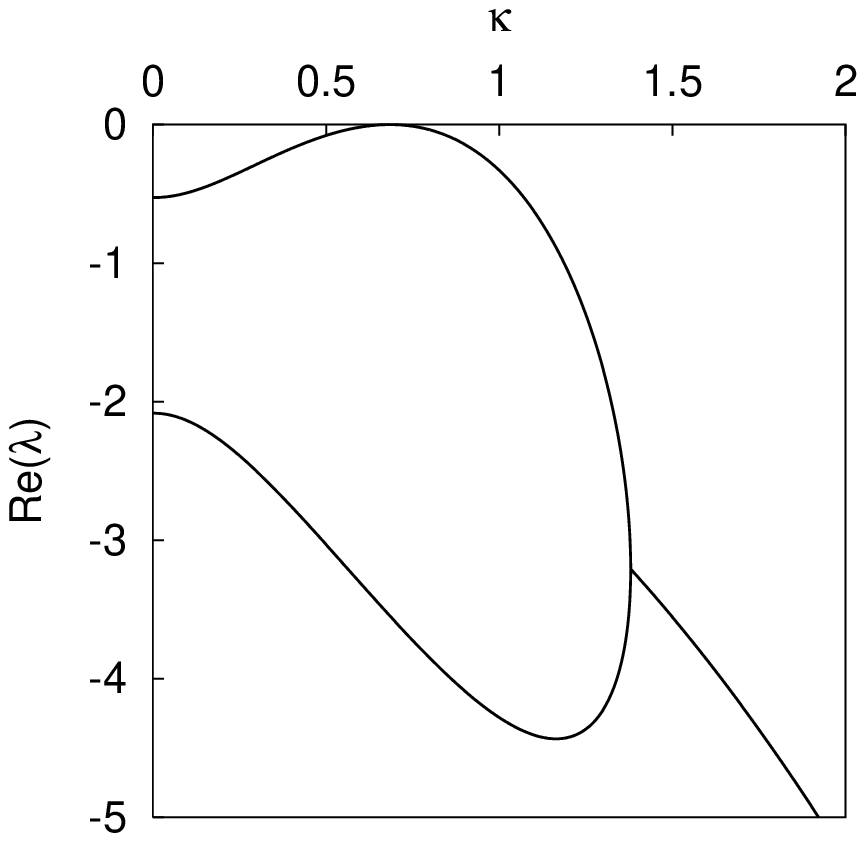}}
\end{picture}
\caption{Growth rate $\mbox{Re}\,\lambda$ of spatially nonuniform
modes as a function of wavenumber $\kappa$ at three different points
on the stability curve displayed in Fig.~\ref{fg:diagram}(a).
The parameters are $\tau=0.29$, $\mu=1.29$ (a),
$\tau=0.36$, $\mu=0.4$ (b), and
$\tau=0.7$, $\mu=0.16$ (c);
other parameters as in Fig. \ref{fg:diagram}.
}
\label{fg:lambdak}
\end{figure}

In Fig.~\ref{fg:lambdak}, we show the real part of $\lambda$ as a
function of $\kappa$ at three different points on the stability curve
in the ($\mu,\tau$) plane. 
The three cases correspond to what has been found earlier for
a different global delayed feedback scheme in the
CGLE~\cite{battogtokh-mikhailov-1996}.
On the branch AB (and similarly also on the branches BC and to the
right of C), the first unstable modes occur with a wavenumber
$\kappa_0\ne0$ where $\mbox{Im}(\lambda)=0$, see
Fig.~\ref{fg:lambdak}(b,c).
Thus, if we cross this branch of the stability boundary by reducing the
feedback intensity below the critical value $\mu_c$, uniform
oscillations will become unstable and standing waves with
wavenumber $\kappa_0$ will emerge (cf. Sec.~\ref{sc:numerics}).
A different situation is encountered on the branch reaching
from A upwards, see Fig.~\ref{fg:lambdak}(a).
Here, the instability will occur by periodic spatiotemporal
modulations of uniform oscillations, since
$\mbox{Im}(\lambda)\ne0$ and the most unstable modes will have
wavenumbers close to $\kappa_0=0$.
%
\section{Numerical simulations}
\label{sc:numerics}
All simulations were carried out for a one-dimensional system of
length $L=128$ on an equidistant grid with $\Delta x=0.32$
(corresponding to a total number of 400 grid points).
We use a second-order finite-difference scheme for the
discretization of the Laplacian operator and impose periodic
boundary conditions.
An explicit Euler scheme with a fixed time step of $\Delta t=0.001$
was employed for integration.
As initial condition, a uniform state superposed with a small
spatially inhomogeneous perturbation was chosen.
The set of parameters is as in the previous section:
$\varepsilon=2$, $\beta=-1.4$, $\omega=2\pi-\beta\approx7.68$,
and $\chi=\pi/2$.
The choice of the feedback parameters $\mu$ and $\tau$ is different
for the different simulations and specified below.

In the first series of simulations, we systematically scan the parameters
$\mu$ and $\tau$ in steps of $\Delta\mu=\Delta\tau=0.1$ between
0 and 2.5, respectively, to verify the shape of the stability domain
in the ($\mu, \tau$) plane.
When starting from uniform initial conditions,
the stability diagram displayed in Fig.~\ref{fg:diagram}(a) is nicely
reproduced after transients.
Asymptotic states are uniform inside the domain of stability and nonuniform
outside this region.
These nonuniform states are, however, of different types.
Far from the region of stability, we asymptotically reach a fully
developed state of defect-mediated turbulence.
As the stability boundary is approached, phase turbulence and
standing wave patterns are observed close to the branches
AB, BC, and to the right of C.

\begin{figure}
\begin{center}
\includegraphics[width=10cm]{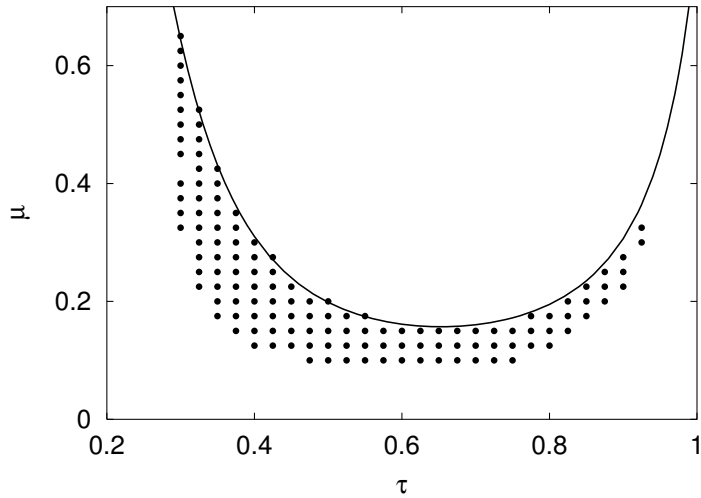}
\end{center}
\caption{Regular spatiotemporal patterns at the border of complete
synchronization for uniform initial conditions.
Section of the ($\mu,\tau$) plane with the analytical result for the
AB branch of the stability curve (cf. Fig \ref{fg:diagram}) plotted
as a solid line.
Bold dots mark the parameters where regular spatiotemporal patters
are observed.}
\label{fg:regular}
\end{figure}

Figure~\ref{fg:regular} shows the results of a more detailed scan
of $\mu$ and $\tau$ in the vicinity of the branch AB of the
synchronization diagram in Fig.~\ref{fg:diagram}(a).
Again, we start from uniform initial conditions.
The feedback parameters $\mu$ and $\tau$ are changed in
steps of $\Delta\mu=\Delta\tau=0.025$ between $\tau=0.25\,...\,0.975$
and $\mu=0\,...\,0.7$.
Simulations that, after transients, resulted in a regular nonuniform
spatiotemporal state are marked by bold dots at the corresponding 
($\mu, \tau$) coordinates
(simulations leading to a uniformly oscillating or turbulent
asymptotic states are not shown).
Obviously, the parameter region, where regular spatiotemporal patterns
occur, constitutes a slightly asymmetric prolongation of the
tongue-shaped stability domain for uniform oscillations towards
smaller feedback intensities.
From the result of the first coarse parameter scan we conjecture, that
the regions where spatiotemporal patterns occur look qualitatively similar
at the other tongue-shaped branches of the stability curve.

\begin{figure}
\begin{center}
\includegraphics[width=10cm]{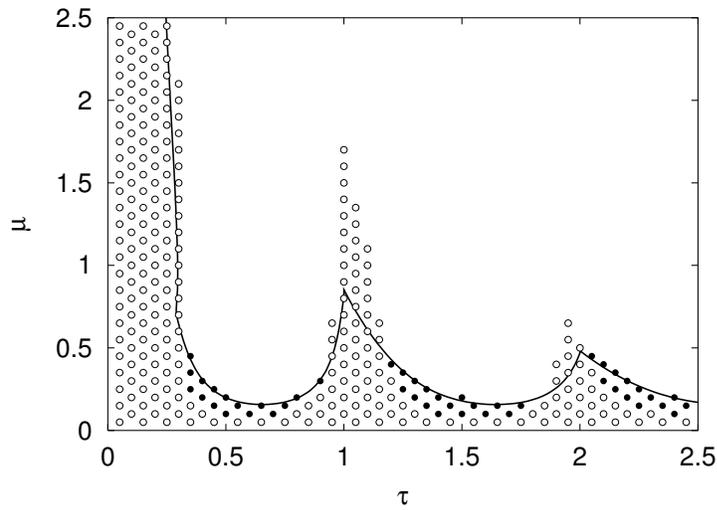}
\end{center}
\caption{Summary of numerical simulations starting from turbulent
initial conditions.
Asymptotic states are shown in the ($\mu,\tau$) plane.
Open circles denote amplitude turbulence and bold circles indicate
regular nonuniform wave patterns.
Symbols are omitted in the areas where computations converge to
uniform oscillations.
The solid line shows the analytically derived stability boundary
for uniform initial conditions.
The parameters are as in Fig.~\ref{fg:diagram}.}
\label{fg:desyn}
\end{figure}

When starting from turbulent initial conditions, the stability
boundary for uniform oscillations is moved towards larger
feedback intensities in the vicinity of the cusps.
The results of numerical simulations are summarized in
Fig.~\ref{fg:desyn}.
Here, open circles indicate a turbulent asymptotic state and bold
circles again denote the appearance of a regular nonuniform wave pattern.
Inside the area where no symbols are shown, simulations
converge towards uniform oscillations.
For comparison, the analytically derived synchronization diagram from
Fig.~\ref{fg:diagram}(a) is also displayed.

\begin{figure}
\setlength{\unitlength}{1mm}
\begin{picture}(140,140)
\put(132,121){(a)}
\put(0,108){\frame{\includegraphics[width=13cm,height=2.5cm]{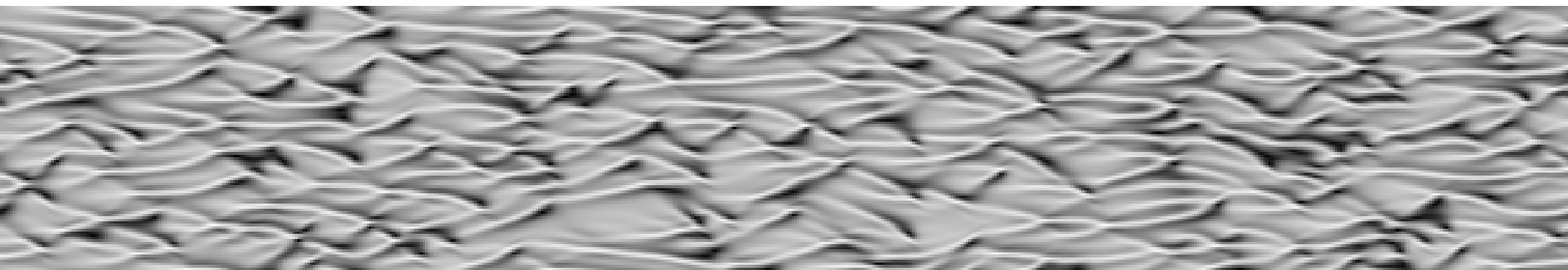}}}
\put(132,94){(b)}
\put(0,81){\frame{\includegraphics[width=13cm,height=2.5cm]{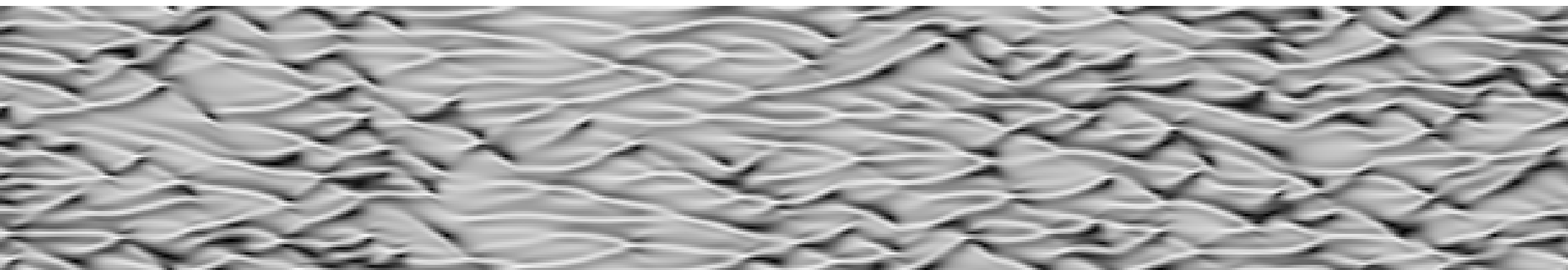}}}
\put(132,67){(c)}
\put(0,54){\frame{\includegraphics[width=13cm,height=2.5cm]{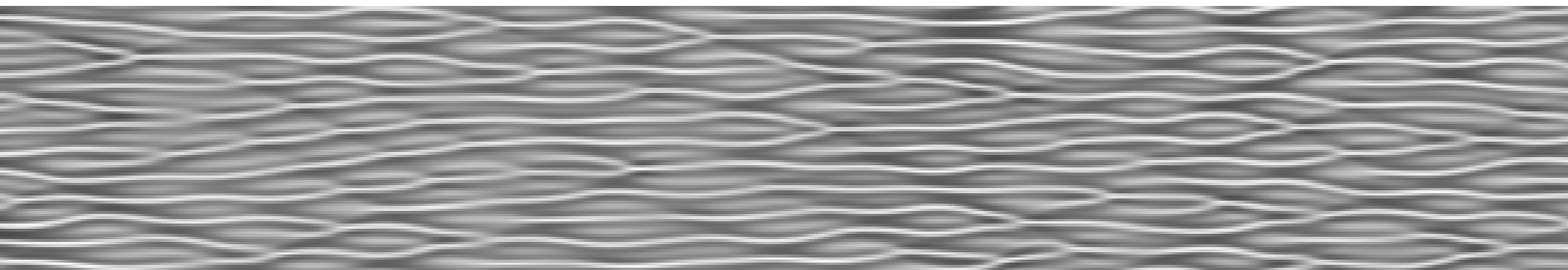}}}
\put(132,40){(d)}
\put(0,27){\frame{\includegraphics[width=13cm,height=2.5cm]{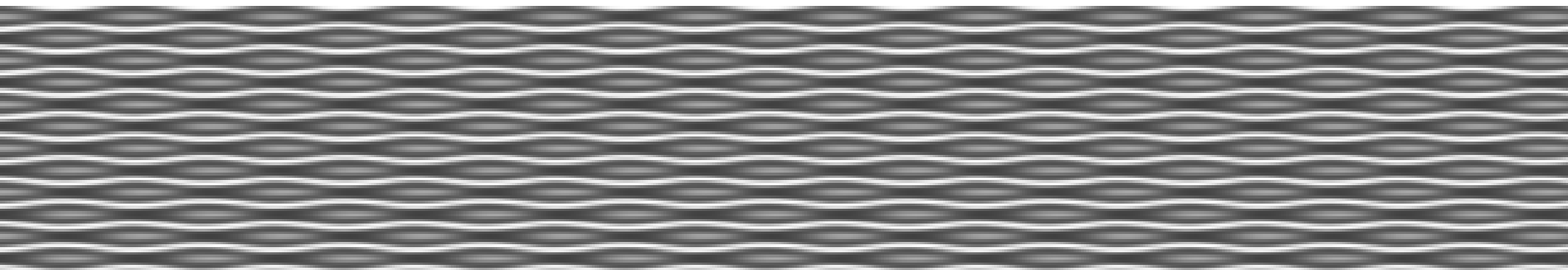}}}
\put(132,13){(e)}
\put(0,0){\frame{\includegraphics[width=13cm,height=2.5cm]{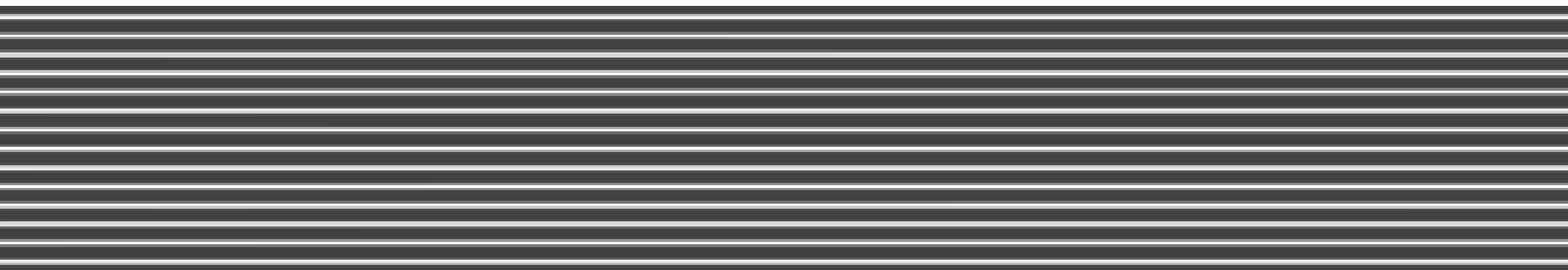}}}
\put(-12,41){space}
\put(-3,-3){\vector(1,0){40}}
\put(40,-4){time}
\put(-3,-3){\vector(0,1){40}}
\end{picture}
\vspace{1mm}
\caption{Spatiotemporal patterns in the parameter region between turbulence
and complete synchronization.
The amplitude $|\eta|$ is displayed in a grey scale coding, where
black (white) denotes low (high) values of the real amplitude.
The delay time is kept constant, $\tau=0.5$ s, and the feedback intensity
increases from top to bottom, (a) $\mu=0$, (b) $\mu=0.05$, (c) $\mu=0.07$,
(d) $\mu=0.1$, and (e) $\mu=0.15$.}
\label{fg:patterns}
\end{figure}

To get a better idea of how the spatiotemporal dynamics of the system
changes in the
parameter range between turbulence and uniform oscillations, we show in
Fig. \ref{fg:patterns} a series of spacetime plots of asymptotic
dynamical states reached for different feedback intensities at a fixed delay
time of $\tau=0.5$.
The amplitude $|\eta|$ is plotted in a grey scale color coding and
the feedback intensity $\mu$ is increased in Fig.~\ref{fg:patterns} from
(a) to (e).
In absence of feedback (Fig.~\ref{fg:patterns}(a), $\mu=0$)
and for small feedback intensities (Fig.~\ref{fg:patterns}(b),
$\mu=0.05$), an irregular state of defect-mediated turbulence is
observed. 
However, the number of defects is smaller in the presence of a weak
feedback and the
time evolution shows intervalls, where almost no defects are seen.
If the feedback intensity is increased
(Fig.~\ref{fg:patterns}(c), $\mu=0.07$), defects are no
longer observed and the system displays a disordered state of phase
turbulence.
For even stronger feedback, breathing
(Fig.~\ref{fg:patterns}(d), $\mu=0.1$) and stationary
standing wave patterns
(Fig.~\ref{fg:patterns}(e), $\mu=0.15$) can be observed.
For still larger feedbacks, $\mu>0.2$ (not shown), uniform oscillations
take place.

To study the behaviour of the 
system for very large delays, a series of simulations with the delays 
$\tau$ varying from 0.5 to 29.5 in steps of $\Delta\tau=1$ has further
been performed.
The feedback intensity $\mu$ was varied for each choice of $\tau$
from 0.1 to 0.3 in the steps of $\Delta\mu =0.1$.
The initial conditions in these 
simulations represented the state of amplitude turbulence reached by 
the system if $\mu=0$.
These numerical investigations have revealed 
that synchronization is possible for all chosen delays, when the 
feedback intensity exceeds a critical level.
In another series of experiments, the feedback intensity was fixed
at $\mu=1$ and the delay time $\tau$ was increased from 0 to 50 in
steps of 0.1.
Here, the transient time was determined for each of the simulations by
looking at the convergence of the statistical variance of the
amplitude $\rho$. 
It was found that for small delays the transient time increases 
proportionally to the delay $\tau$.
For large delays, the transient time undergoes saturation at a level
independent of $\tau$, but depending on the intensity of the applied
feedback.

For very long delays $\tau$, the 
component $\bar{\eta}(t-\tau)$ in the feedback signal $F(t)$ at 
time $t$ corresponds to the initial state of amplitude turbulence. 
Thus, the global delayed feedback scheme in the considered 
limit effectively represents global forcing of the system with an 
external chaotic signal corresponding to the initial turbulent state in 
absence of feedback.
To verify this conjecture, special numerical simulations have been
performed.
In these simulations, the feedback 
signal was generated by replacing $\bar{\eta}(t-\tau)$ with the 
average complex oscillation amplitude of the 
same system without the feedback.
We have found that, by applying such chaotic external forcing,
synchronization of uniform oscillations can also be achieved.
%
\section{Discussion}
\label{sc:discussion}
Motivated by our experiments~\cite{beta-et-al-2003} on control of chemical
turbulence in the CO oxidation 
reaction on Pt(110), we have performed a detailed study of the effects of 
time-delay autosynchronization on uniform oscillations in a general 
model described by the complex Ginzburg-Landau equation.
We have seen 
that, like in the experiments, the noninvasive stabilization of uniform 
oscillations by this method is not possible, though the required 
magnitude of the feedback signal can be significantly reduced by using 
an optimal delay time.
The obtained synchronization diagram exhibits a 
series of cusps at the values of delay times equal to integer multiples 
of the oscillation period of the unperturbed uniform system.
This feature seems to be common for various oscillatory systems with delayed 
coupling and has also been found in the case of the Kuramoto model of 
phase oscillators with a delayed global coupling~\cite{yeung-strogatz-1999}.
Near the 
synchronization boundary, the formation of standing 
waves and a state of modified intermittent turbulence were 
observed.
These effects are similar to those previously discussed for a 
different feedback scheme~\cite{bertram-mikhailov-2003,battogtokh-mikhailov-1996,battogtokh-preusser-mikhailov-1997}.
In contrast to our previous studies, the action of feedbacks with long
delay times has also been investigated.
We have shown that such feedbacks are also capable 
of synchronizing oscillations and this effect can essentially be 
explained as suppression of turbulence by external noisy forcing.
%
%

%
\newpage
\section{Figures}
\label{sc:fig}
\vspace{10mm}
%
\end{document}